\documentstyle[twoside,fleqn,espcrc2]{article}

\newcommand{\eq}{\begin{equation}}
\newcommand{\en}{\end{equation}}
\newcommand{\eqa}{\begin{eqnarray}}
\newcommand{\ena}{\end{eqnarray}}
\newcommand{\noi}{\noindent}
\newcommand{\np}{n^{\prime}}
\newcommand{\npp}{n^{\prime \prime}}
\newcommand{\spr}{s^{\prime}}
\newcommand{\sprr}{s^{\prime \prime}}
\newcommand{\xpr}{x^{\prime}}
\newcommand{\xprr}{x^{\prime \prime}}
\newcommand{\smgr}{\stackrel{\textstyle <}{>}}
\newcommand{\grsm}{\stackrel{\textstyle >}{<}}
\newcommand{\aleq}{\mbox{}_{\textstyle \sim}^{\textstyle < }}
\newcommand{\ageq}{\mbox{}_{\textstyle \sim}^{\textstyle > }}
\newcommand{\ra}{\rightarrow}
\newcommand{\lra}{\longrightarrow}
\newcommand{\gm}{\gamma_{\mu}}
\newcommand{\gn}{\gamma_{\nu}}

\newcommand{\AmS}{{\protect\the\textfont2
  A\kern-.1667em\lower.5ex\hbox{M}\kern-.125emS}}

\title{
\vspace{-1.8cm}
\hbox{}
{\small NOVEMBER 1993} \hfill {\small DESY 93--180~~}           \break
\hbox{}                \hfill {\small HU BERLIN--IEP--93/6~~}   \break
                                                              \break
Phase structure and chiral limit of compact lattice QED
with Wilson fermions
\thanks{TALK GIVEN AT THE LATTICE '93 INTERNATIONAL SYMPOSIUM
       LATTICE FIELD THEORY, DALLAS, USA, OCTOBER 12--16, 1993}
\thanks{Work supported by the Deutsche
Forschungsgemeinschaft under research grant Mu 932/1-1 }
}

\author{A. Hoferichter\address{Fachbereich Physik,
Humboldt-Universit\"at zu Berlin, 10099 Berlin, Germany},
V.K.~Mitrjushkin$\mbox{}^{\mbox{\scriptsize a}}$
\thanks{Permanent address: Joint
Institute for Nuclear Research, Dubna, Russia},
M.~M\"uller-Preussker$\mbox{}^{\mbox{\scriptsize a}}$
and Th.~Neuhaus\address{Fakult\"at f\"ur Physik,
Universit\"at Bielefeld, 33615 Bielefeld, Germany}
}

\begin{document}

\begin{abstract}
We study the phase structure and chiral limit of $4d$ compact
lattice QED with Wilson fermions (both dynamical and quenched).
We use the standard Wilson action (WA) and also the modified action
(MA) with some lattice artifacts suppressed.
We show that lattice artifacts influence the distributions of
eigenvalues $~\lambda_i~$ of the fermionic matrix especially for
small values of $~\lambda_i~$. Our main conclusion is that the
chiral limit of compact QED can be efficiently located
using different techniques.
\end{abstract}

\maketitle

\section{Introduction}


The lattice formulation of QED is not unique.
One has to decide on a physical ground which version of QED is
realised in nature if different lattice versions of QED do not
belong to the same universality class.
The {\it old} physics, i.e., known from experiment, has to be
reproduced.

When we consider QED as arising from a subgroup of some non--abelian
(e.g., grand unified)
gauge theory we are necessarily dealing with the compact version.
Our choice in this work is a {\it compact} formulation of QED.

In the theory with Wilson's fermions \cite{wil} chiral symmetry is broken
explicitly, and, presumably, can be only restored by fine-tuning
the parameters in the continuum limit if we are dealing with
a meaningful lattice discretization.
What one can expect at nonzero spacing is that at some
$\kappa_{c} \equiv \kappa_{c}(\beta )$ a so--called {\it partial}
symmetry restoration takes place \cite{kawa,kasm} when the Wilson
mass term and ordinary mass term cancel at zero momentum
in certain vertex functions. If so one can
approach continuum limit and chiral symmetry restoration
along the line $\kappa_{c}(\beta )$.
Another question is if in the continuum limit of our theory
the chiral symmetry is realised explicitly or is spontaneously broken.


\section{Actions and order parameters}

The modified lattice action
$ S_{MA}(U, {\bar \psi}, \psi)$ for $4d$ $~U(1)~$ gauge theory (QED) is

\eq
 S_{MA} = \beta \cdot S_{G}(U) + S_{F}(U, {\bar \psi}, \psi) +
\delta S_{G}(U) .
                                              \label{ma}
\en

\noi In eq.(\ref{ma}) $~S_{G}(U)~$ is the standard plaquette
(Wilson) action for the pure gauge $~U(1)~$ theory, and the
additional term $~\delta S_{G}~$ suppresses
lattice artifacts (i.e., monopoles and negative plaquettes).

\noi The fermionic part of the action
$S_{F}(U, {\bar \psi}, \psi)~$ is

\vspace{-0.6cm}
\eqa
S_{F}
& = & \sum_{f=1}^{N_{f}} \sum_{x,y} \sum_{s,\spr=1}^{4}
{\bar \psi}_{x}^{f,s} {\cal M}^{s \spr}_{xy} \psi_{y}^{f,\spr}
\equiv  {\bar \psi} {\cal M} \psi~,
\nonumber \\
{\cal M} & \equiv & \hat{1} + \kappa \cdot \tilde{{\cal M}}(U) ,
                                              \label{waf}
\ena

\vspace{-0.1cm}
\noi where $~{\cal M}~$ is Wilson's fermionic matrix,
$~N_{f}~$ is the number of flavours and $\kappa$ is the hopping parameter.
The first two terms in r.h.s of eq.(\ref{ma}) make up the standard
Wilson action $S_{WA}$.

In our work we used both $S_{WA}$ and $S_{MA}$ with
$N_f=2$ for dynamical fermions.
Apart from $\langle {\bar \psi} \psi \rangle$ and
$\langle {\bar \psi} \gamma_{5}  \psi \rangle$ we calculated
the {\it pion norm} $\langle \Pi \rangle$ \cite{bkr}

\eq
\Pi (U) = \frac{1}{L^4} \cdot
\sum_{x} \mbox{Tr} \Bigl( {\cal M}^{-1}_{~0x} \gamma_{5}
{\cal M}^{-1}_{~x0} \gamma_{5} \Bigr) ,
                                      \label{fop}
\en

\noi where $L$ is the lattice size.
Its advantage is that it appears to be a very sensitive
quantity in the 'critical' region.
Introducing  eigenvectors $g_{n}(s,x)$ of
$~\gamma_{5} {\cal M}~$ with eigenvalues $~\mu_{n}$:
$~\gamma_{5} {\cal M} g_{n} = \mu_{n} \cdot g_{n}~$,
one can easily obtain a spectral representation

\eq
\Pi = \frac{1}{L^4} \sum_{n} \frac{1}{\mu_{n}^2}
\sum_{s} \mid \! g_{n}(s,0) \! \mid^2~.
\en

\vspace{-0.2cm}
\noi Following the common practice we identify here the chiral
transition with the appearance of zero or near to zero eigenvalues of the
fermionic matrix ${\cal M}$.
 Evidently, an eigenstate of ${\cal M}$ with eigenvalue zero is
also an eigenstate of $\gamma_{5} {\cal M}$. So, the presence of
configurations which belong to zero eigenvalues of $~{\cal M}~$
gives rise to poles in $~\Pi~$.

%
%

\section{Conjugate gradient (cg--) method}

To locate $\kappa_{c}(\beta)$ one can use the convergence
rate of the cg--method. This is the iterative method of
solving a system of linear equations $D \cdot X = \varphi $,
where $D$ is a hermitian $n \times n$ matrix and
$\varphi$ is an input vector ($D = {\cal M}^{\dagger}{\cal M}$
in our case).
The convergence of the cg--method should be controlled by the condition
number $~\xi \equiv \lambda_{max}/\lambda_{min}~$,
 where $~\lambda_{max}~$ and $~\lambda_{min}~$ are maximal and
minimal eigenvalues of $D$.
 Close to $~\kappa_{c}~$ the minimal eigenvalue of
$~{\cal M}^{\dagger} {\cal M}~$ is small and is  supposed to be
$~\sim \bigl( 1 - \kappa/\kappa_{c} \bigr)^2~$.

We observed that at large enough number of iteration steps
$~N_{cg}~$ ($N_{cg} > N_{0}$) the average residue
$~\langle R \rangle \equiv \langle R \rangle (N_{cg})~$ behaves as

\eq
\langle R \rangle = C \cdot \exp (-\alpha \cdot N_{cg} ),
\quad \alpha = \ln \frac{\sqrt{\xi}+1}{\sqrt{\xi}-1}
\en

\noi {\it independently of the distribution of eigenvalues $\lambda_{i}$}
provided $~n~$ is large enough.
To check it we generated $~D~$ with different
(uniform, gaussian, double--peaked) distributions
of eigenvalues and given $~\lambda_{min}~$ and $~\lambda_{max}~$.
The components of the initial vector $~X_{0}~$ and of
$\varphi$ were chosen every time randomly with gaussian distributions.

%
%
%

It follows from the above that, for the inversion of
$~{\cal M}^{\dagger} {\cal M}~$, the
$\langle N^{-1}_{cg} \rangle$ required for
convergence to some small but fixed $R$ will behave as
$\langle N^{-1}_{cg} \rangle \sim 1 - \kappa/\kappa_{c}$.
In Fig.1 we show the dependence of
$\langle N^{-1}_{cg} \rangle$ on $~\kappa~$ at $~\beta = 0.8~$
for quenched MA (qMA) for different $L$.

\begin{figure}[htb]
\vskip 52mm
\caption{$\langle N^{-1}_{cg} \rangle(\kappa)$ for qMA for different $L$.}
\label{fig:cg}
\end{figure}

At $\kappa \sim \kappa_c$ the data fit nicely to straight lines
giving reasonable estimation of $\kappa_c$. The volume
dependence
becomes rather weak  for larger $L$.
A similar picture was obtained for the case of dynamical fermions.

\section{Phase diagram and chiral limit}


For WA with dynamical fermions (dWA) at
$\beta < \beta_0 \sim 1.0$
thermal cycles with respect to $\kappa$ or $\beta$ have a typical
hysteresis behaviour for ${\bar \psi} \psi $ and plaquette $\Box $.
Time histories (TH's) of ${\bar \psi} \psi $ and plaquette $\Box $ for
different starts show the existence of metastable states.
So, we conclude that for dWA there is a $1$st order phase
transition (PT) line from $(\beta;\kappa) \simeq (1.;0.)$
to $(\beta;\kappa) = (0.;0.25)$ which is in agreement with \cite{hege}.

After the suppression of lattice artifacts (i.e., for dMA)
this line disappears (see also \cite{bmmp}).

At $\beta > \beta_0$ and $\kappa < \kappa_c(\beta)$ the system
with dWA is in the Coulomb phase. The photon correlator
$\Gamma(\tau)$ is well
consistent with that corresponding to a zero photon mass.
For $\beta > \beta_0$ and $\kappa > \kappa_c(\beta)$
the correlator $~\Gamma(\tau)$ shows a tachyonic--type behaviour
($m^2_{\gamma} < 0$).
Thus, we can conclude that there is a
higher--$\kappa$ phase (or phases) differing from the Coulomb phase.


For quenched WA (qWA) at $\beta < \beta_0$
TH's of $\Pi$ (as well as of $\overline{\psi} \psi$ and
$\overline{\psi} \gamma_{5} \psi$) show very sharp peaks
at $\kappa \sim \kappa_c(\beta)$ which means the appearance
of small eigenvalues $\lambda_i$.
It is worth noting that for WA at $\beta < \beta_0$
those peaks do {\it not} disappear at $~\kappa > \kappa_c~$
but instead become even more strong.

For WA at $~\beta > \beta_0$  and for MA at any (positive)
$\beta$ the dependence of TH's on $\kappa$ changes drastically.
We don't find peaks of comparable amplitude ($\sim 10^4$) but,
nevertheless, in some 'critical' region  $\kappa \sim \kappa_c$
TH's of $~\Pi~$ become much more rough than at smaller or {\it larger}
values of $\kappa$. As far as for
every configuration $~\Pi~$ is the arithmetic average of $~4 L^4~$
terms corresponding to $~4 L^4~$ different eigenvalues
one can conclude that rather small $~\lambda_i~$ appeared.

The (renormalized) variance of the pion norm
$\sigma^2(\Pi) \equiv L^4 \cdot \mbox{Var}(\Pi)$ appears to be
a suitable 'order parameter'.

\begin{figure}[htb]
\vskip 52mm
\caption{$\sigma^2(\Pi)$ for qMA at $\beta =0.8$.}
\label{fig:pinorm}
\end{figure}

 Fig.2 shows $\sigma^2(\Pi)$ for qMA at $\beta=0.8$.
There is  a clear signal at some $\kappa_c$.
With increasing $L$ this peak becomes even more sharp.

For dynamical fermions the signal becomes less sharp
(at least at small $L$) because the fermionic determinant
suppresses small eigenvalues.
However also in this case $\sigma^2(\Pi)$ develops a pronounced
maximum thus allowing to locate $\kappa_c$.

In Fig.3 we show the phase diagrams for both dWA and dMA.
At the moment we have no clear interpretation of the
phase in the upper left corner in Fig.3a.

\begin{figure}[htb]
\vskip 50mm
\caption{Phase diagrams for WA(a) and MA(b).}
\label{fig:phased}
\end{figure}

\vspace{-0.6cm}
\section{Conclusions}

For the standard compact Wilson action we observe a presumably
$1$st order PT which disappears after suppressing lattice artifacts.

These lattice artifacts influence strongly the  distribution
of eigenvalues $~\lambda_{i}~$ of  the fermionic matrix ${\cal M}$.
This influence is especially pronounced for the near--to--zero
values of $\lambda_{i}$.

 After suppression of artifacts a chiral
transition (i.e., appearance of near--to--zero eigenvalues
of ${\cal M}$) is left on a 'horizontal' line
$\kappa = \kappa_c(\beta)$.


As a preliminary conclusion, we have {\it no} sign
for a qualitative change of the behaviour along
the line $\kappa = \kappa_c(\beta)$.

\end{document}